# $Nb_xTi_{1-x}N$ low timing jitter single-photon detectors with unity internal efficiency at 1550 nm and 2.5 K

**Julien Zichi,**[1,*] **Jin Chang,**[2] **Stephan Steinhauer**[1], **Kristina von Fieandt,**[3] **Johannes W. N. Los,**[4] **Gijs Visser,**[4] **Nima Kalhor,**[4] **Thomas Lettner,**[1] **Ali W. Elshaari,**[1] **Iman Esmaeil Zadeh,**[2] **and Val Zwiller**[1,4]

[1]*Department of Applied Physics, Royal Institute of Technology (KTH), SE-106 91 Stockholm, Sweden*
[2]*Optics Research Group, ImPhys Department, Faculty of Applied Sciences, Delft University of Technology, Lorentzweg 1, 2628 CJ Delft, The Netherlands*
[3]*Inorganic Chemistry Research Programme, Department of Chemistry – Ångström Laboratory, Uppsala University, Box 538, SE-751 21 Uppsala, Sweden*
[4]*Single Quantum B.V., 2628 CH Delft, The Netherlands*
*\*zichi@kth.se*

**Abstract:** The requirements in quantum optics experiments for high single photon detection efficiency, low timing jitter, low dark count rate and short dead time have been fulfilled with the development of superconducting nanowire single photon detectors. Although they offer a detection efficiency above 90%, achieving a high time resolution in devices made of amorphous materials is a challenge, particularly at temperatures above 0.8 K. Devices made from niobium nitride and niobium titanium nitride allow to reach the best timing jitter, but in turn have stronger requirements in terms of film quality to achieve a high efficiency. Here we take advantage of the flexibility of reactive co-sputter deposition to tailor the composition of $Nb_xTi_{1-x}N$ superconducting films, and show that a Nb fraction of x = 0.62 allows for the fabrication of detectors from films as thick as 9 nm and covering an active area of 20 μm, with a wide detection saturation plateau at telecom wavelengths and in particular at 1550 nm. This is a signature of an internal detection efficiency saturation, achieved while maintaining the high time resolution associated with NbTiN and operation at 2.5K. With our optimized recipe, we reliably fabricated detectors with high critical current densities reaching a saturation plateau at 1550 nm with 80% system detection efficiency, and with a FWHM timing jitter as low as 19.47 ps.

## 1. Introduction

The invention [1] and active development of superconducting nanowire single photon detectors (SNSPDs) have offered in recent years a unique tool to study and harvest the physics of single photons, from fundamental quantum optics experiments to impactful applications. Characterization of single photon emitters such as quantum dots [2] or color centers [3], loophole-free Bell test measurements [4], laser ranging [5,6], quantum-secure communication [7,8], CMOS testing [9,10] or biological imaging [11] all take advantage of the unique combination of sensitivity, broad wavelength range, negligible dark count rate, and excellent time resolution of SNSPDs. The demonstration of SNSPDs based on amorphous WSi achieving 93% system detection efficiency [12] opened the way for the fabrication of high-yield, high efficiency detectors. However, the operation at sub-Kelvin temperature of these detectors, combined with a typically high timing jitter [12,13] limits their range of applications. On the other hand, NbN and NbTiN detectors have shown a combination of high efficiency above 90%, low jitter of 10 ps and below and low dark count rate, together with a maximum count rate of hundreds of MHz and with a typical operation temperature > 2 K [14–17] but the fabrication of detectors with a good internal detection saturation at 1550 nm in these materials remains a challenge, requiring high quality films, lithography and etching [18]. Overall, this limits the fabrication yield of high performance detectors. Here we take advantage of the extra degree of freedom offered by the co-sputtering deposition of NbTiN to tailor the Nb content during the film deposition on standard $SiO_2$/Si substrates. The critical temperature ($T_c$) of the films was measured, and we carried out X-Ray Photoelectron Spectroscopy (XPS) analysis to precisely measure the concentration of niobium, titanium, and nitrogen. In parallel, we deposited selected NbTiN recipes on 10 nm thin silicon nitride support films and characterized their crystalline structure and nanoscale morphology with transmission electron microscopy (TEM). Finally, we patterned the films into SNSPDs, and measured the photon count rates of the detectors as a function of applied bias in a Gifford-McMahon cryostat operating at 2.5 K, showing that our optimized film composition allows for the fabrication of devices with a pronounced detection efficiency saturation at 1310 nm and 1550 nm. The improved recipe was then used to deposit a film on a Distributed Bragg Reflector (DBR) microcavity, and detectors were fabricated and fully packaged using a standard self-aligning method to carefully measure their system detection efficiency (SDE), dark count rate (DCR) and timing jitter.

## 2. Superconducting thin film deposition and characterization

The superconducting thin films were deposited by reactive magnetron co-sputtering from two separate targets of Nb and Ti, in an atmosphere of Ar and $N_2$, as depicted in Fig. 1.(a), on dry thermally oxidized silicon wafers. We deposited another set of films on 10 nm-thick SiN TEM support films, for 4 different recipes spanning the range of compositions investigated. The substrates were placed on a sample holder that was rotated during the depositions to maximize the films homogeneity. The base pressure of the system was kept between $5\times10^{-9}$ and $1\times10^{-8}$ Torr before the deposition; a mixture of 100 sccm Ar and 10 sccm $N_2$ was introduced and the pressure was set to 28 mTorr in order to strike the plasmas with 50 W for each target. The pressure was lowered to 3 mTorr while the plasma powers were ramped up to their targeted values. For our main series of films from which detectors were fabricated, the Ti target was biased with 240 W RF, while the Nb target powers were set from 240 W DC down to 60 W DC, in 20 W steps. A substrate shutter and a rate monitor were used to precisely control the deposition time and to determine the deposited thickness. For each recipe, we first deposited a film with a thickness in the order of 50 nm measured with a stylus profilometer, which we used to calibrate the deposition rate of each recipe, to measure the film $T_c$ of the material, and to carry out XPS measurements. Subsequently, we used the deposition rate calibration to grow films with a nominal thickness of 9 nm (± 5%), which were used to measure the thin film room temperature sheet resistance, the thin film $T_c$, and which were eventually used to pattern the SNSPDs.

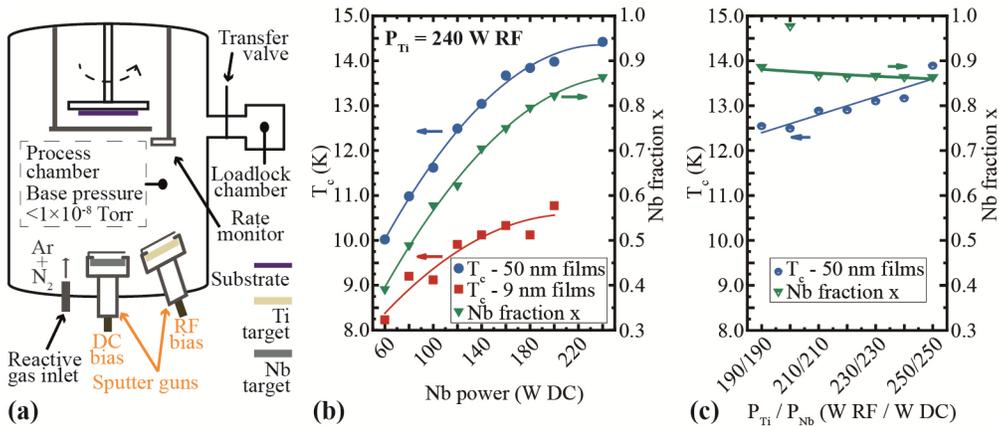

Fig. 1. (a) Schematics of the sputtering chamber. (b) Critical temperature $T_c$ and Nb fraction x in the alloy $Nb_xTi_{1-x}N$ vs. Nb sputtering power. The power applied to the Ti target was kept constant at 240 W RF for all film depositions. (c) Critical temperature $T_c$ and Nb fraction x vs. sputtering power for a constant power ratio. The lines are a guide to the eye.

For electrical characterization, at least two samples of each film were carefully diced into squares and placed into a custom-made four-point $T_c$ measurement setup which consists of a closed-cycle cryostat built around a cold head with a nominal base temperature of 4.2 K. The samples were contacted with a linear four-pin probe with spring-loaded pins, while the temperature was ramped up and down at a rate of 0.1 K/min. We measured the room temperature sheet resistance and the $T_c$, defined as the temperature at which the resistivity of the film was half of that at 20 K.

In parallel, XPS measurements were performed to study the different compositions of the 50 nm films, using a PHI Quantera II Scanning ESCA microprobe with monochromatic Al Kα radiation and a spot size of 100 µm. The composition of the top surface, a native oxide layer as observed elsewhere [19], as well as the average bulk composition of the superconducting layer were determined. For the latter, 1 kV $Ar^+$ sputter etched depth profiles with an area of 1x1 $mm^2$ were conducted, the thickness of 50 nm allowing for the etching without removing the underlying NbTiN layer and for a comparing values close to bulk conditions. The sensitivity factors for determining the composition were obtained by reference measurements using time-of-flight elastic recoil detection analysis [20]. The XPS spectra reveal a composition of the superconducting layer without contaminant species and with a content of N of 50 ± 3 %, which hints that our films can be considered as a mixture of stoichiometric NbN and TiN [21] where the proportions of Nb and Ti depend on the sputtering parameters. In Fig. 1.(b) we compile the results from the $T_c$ and the XPS measurements and show that we can precisely control the composition of the films in terms of Nb fraction x in the alloy $Nb_xTi_{1-x}N$ with a value ranging widely from 0.39 to 0.86. We confirm that films with a higher Nb content yield a higher $T_c$, varying from 10.02 K to 14.42 K for 50 nm films. To decouple the effect of the total sputtering power on the film growth and on the $T_c$ of the material, we deposited another series of films, keeping the power ratio on both targets constant from $P_{Ti} = P_{Nb} = 190$ W, up to $P_{Ti} = P_{Nb} = 250$ W. We observed a much lower impact of the total power density on the $T_c$ and a negligible impact on the Nb fraction x (Fig. 1.(c)). We attribute the slight shift of the $T_c$ from 12.5 K to 13.5 K to a change in the film density of our material, which is confirmed by the observation of lower deposition rates for higher sputtering power. We conclude that the effects we report in this article are mainly due to a change in the chemical composition of the material.

We controlled the thickness of the films by measuring lift-off steps with an atomic force microscope (AFM), as shown in Fig. 2.(a). The roughness of a selection of films was extracted from the AFM measurements, and we observed root mean square (Rq) roughness values between 0.7 and 0.9 nm. Furthermore we used TEM to study the crystalline structure and nanoscale morphology of NbTiN obtained with three different recipes (Nb powers of 60 W, 120 W and 180 W; constant Ti power of 240 W RF). The instrument was operated at electron acceleration voltage of 300 kV and was equipped with an aberration corrector for the objective lens. The high resolution TEM (HRTEM) image presented in Fig. 2.(b) was taken on the film with $P_{Nb}$ = 120 W (Nb fraction x = 0.62) and reveals that our material is polycrystalline. The inset shows the Fourier transform of a single grain, which was indexed to the [110] zone axis of the face-centered cubic structure of NbTiN. For the three investigated samples, there was no sizeable change in the dimension and distribution of grains, which have on average a diameter of 4 to 5 nm, indicating that the growth mode of the different films was not affected when modifying the recipes. In Fig. 2.(c) selected area electron diffraction results are presented, showing a ring pattern typical for polycrystalline films. As can be seen in the inset, the [200] lattice spacing (obtained by azimuthal integration using the PASAD plug-in [22] and Digital Micrograph software) was found to scale with the Nb fraction x, which is expected for the fully miscible ternary NbTiN alloy [19]. We note that the lattice parameters were extracted from films grown on SiN TEM support films with a thickness of 10 nm, which most probably does not correspond to the lattice parameters of films deposited on $SiO_2$/Si substrates due to different levels of microscopic strain [23]. To assess the influence of the substrate material in general, we directly compared the superconducting properties of two samples, one on SiN and one on $SiO_2$, deposited during the same sputtering run. We measured a difference in $T_c$ of only 0.07 K for the recipe with $P_{Nb}$ = 120 W, which indicates that the material growth modes on both substrates are similar. Hence, it is expected that the results presented here are relevant for a broader range of silicon photonics platforms, e.g. integrated circuits relying on SiN-based waveguides [24,25].

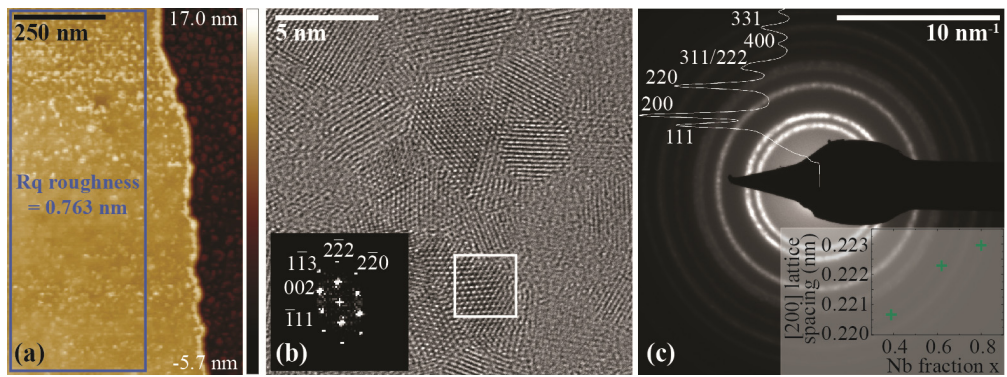

Fig. 2. Characterization of films with an Nb fraction x = 0.62. (a) AFM map showing the thickness measurement step, and the Rq roughness value of 0.763 nm extracted from the region of interest. (b) HRTEM micrograph of NbTiN deposited on a SiN support film. Inset: Fourier transform of a single grain (indicated by the white square on the micrograph) imaged along the [110] zone axis. (c) Selected area electron diffraction of the film. Inset: the extracted [200] lattice spacing scaled with Nb fraction x for different samples.

## 3. Detectors

The binding energy of the Cooper pairs in the superconductor as described in the Bardeen-Cooper-Schrieffer theory [26] and in the Ginzburg-Landau description of the SNSPD model [27] dictates the $T_c$ of the films of a fixed thickness. It influences the amount of quasi-particles that can be created upon absorption of a single photon of any given energy [28]; as a result, the $T_c$ strongly affects the intrinsic detection efficiency of an SNSPD made from that material. Here we apply this concept and verify the correlation between the $T_c$ of our material and the intrinsic detection efficiency of detectors made from eight different recipes and material compositions. Fig. 3.(a) depicts an SEM image of a meandering SNSPD patterned from one of the films, following the fabrication steps described by the authors elsewhere [14]. The meander covers a 10 μm of diameter area, with a 50% filling factor and a line width of 70 nm. A total of sixteen detectors were fabricated from each film, and six were tested at 2.5 K in a closed-cycle cryostat. The structures were flood-illuminated with continuous-wave lasers at two different wavelengths – 1310 nm and 1550 nm – and their response to light was characterized as a function of bias current.

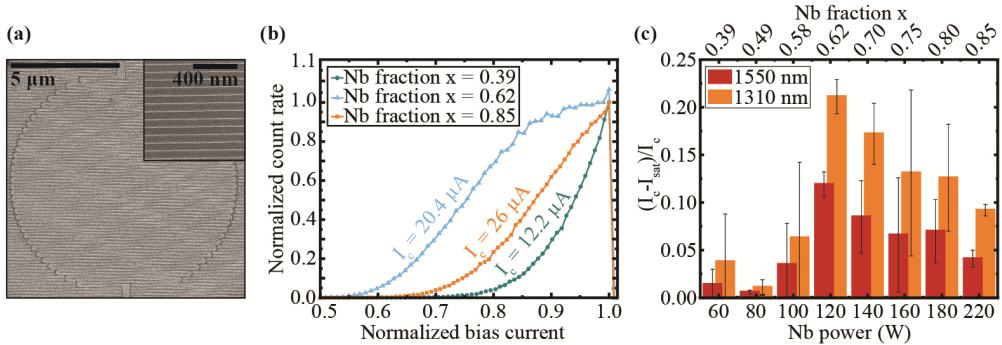

Fig. 3. (a) SEM micrograph of the SNSPD structures patterned from the films. Inset: close-up view on the meander nanowire lines. (b) Examples of normalized count rate at 1550 nm vs normalized bias current of representative detectors from three different film recipes, all with a thickness of 9 nm. (c) Average relative saturation plateau ($I_c$-$I_{sat}$)/$I_c$.

In Fig. 3.(b) we show the normalized count rates of the detectors with the highest $I_c$ from three different film recipes against the normalized bias current, and we observe different shapes of the photon count rate curves. For films with an Nb content of x = 0.39 and x = 0.85, the curves have an exponential shape up to the critical current, whereas for the film with x = 0.62 a saturation plateau appears at around 80% of the $I_c$. In Fig. 3.(c) we show the average of the relative saturation plateau width of the devices made from each deposition recipe, which we define as the difference between the critical current $I_c$ and the saturation current $I_{sat}$ normalized to the critical current $(I_c-I_{sat})/I_c$, and where $I_{sat}$ is the bias current value at which the count rate is equal to 90% of its maximum value. We observe an optimum of this plateau width for the recipe with x = 0.62 that can be explained by the fact that for a higher Nb content the higher $T_c$ implies a stronger binding of Cooper pairs in the material, reducing the internal detection efficiency. On the other hand, for lower Nb contents, we observe that the critical current is reduced, although it remains comparatively high for the operation temperature of 2.5K due to the relatively large thickness of the films. By reducing the Nb fraction in the alloy to 0.62 as measured by XPS, we can fabricate detectors with large cross-sections of 9 nm × 70 nm with a saturated detection efficiency at 2.5 K, a large critical current and a high yield with six working detectors out of six tested and with a narrow spread of performances, due to a lesser sensitivity to the substrate surface roughness. The parameters extracted from the films and detectors made for each recipe are summarized in Table 1.

**Table 1: Summary of films and corresponding detectors parameters**

| | Films | | | | Detectors | |
|---|---|---|---|---|---|---|
| Nb fraction x | Nb power[1] W | Tc – 50 nm films K | RT sheet resistance Ω/□ | Residual resistance ratio – 9 nm films | Max. $J_c$ ×10$^9$ A/m$^2$ | Average relative saturation plateau width at 1550 nm |
| 0.86 | 240 | 14.42 | No film | No film | No detector | No detector |
| 0.85[2] | 220 | No film | 480.25 | 3.80 | 20.79 | 0.04 |
| 0.82 | 200 | 13.98 | 445.40 | 0.91 | No detector | No detector |
| 0.80 | 180 | 13.84 | 598.54 | 0.79 | 16.83 | 0.07 |
| 0.75 | 160 | 13.67 | 539.13 | 0.80 | 14.52 | 0.07 |
| 0.70 | 140 | 13.04 | 595.02 | 0.81 | 15.79 | 0.09 |
| 0.62 | 120 | 12.49 | 320.56 | 0.41 | 16.98 | 0.12 |
| 0.58 | 100 | 11.62 | 550.83 | 0.26 | 12.46 | 0.04 |
| 0.49 | 80 | 10.98 | 608.39 | 0.88 | 9.68 | 0.01 |
| 0.39 | 60 | 10.02 | 658.35 | 0.92 | 9.76 | 0.01 |

[1]Ti sputter power constant at 240 W RF.
[2]From fit.

After selecting the optimal material recipe, we deposited a film with a Nb content x = 0.62 on a substrate consisting of a DBR micro-cavity designed to enhance light absorption at the surface for a wavelength of 1550 nm. We packaged detectors into fiber-coupled chips using a self-aligning technique [29], and tested them in a 2.5 K cryostat, allowing us to precisely measure the system detection efficiency and timing jitter. We kept the same filling factor and linewidth of the nanowire, and designed detectors with a 20 µm active area which is suitable for coupling not only with single mode fibers at 1550 nm but also with graded index multi-mode fibers, facilitating the signal coupling into the fiber for a broad range of optical instruments. The DCR only becomes noticeable at a bias current of around 98% of the $I_c$, with some background counts present at lower bias accounting for < 200 counts per second, which we attribute to a residual coupling of black-body radiation, as presented for one of the detectors in Fig. 4.(a). The use of fibers with DBR filtering on their end-facet as proposed elsewhere [30] is identified as a solution for reducing the background noise further. Using a diode laser operating at 1550 nm, a calibrated power meter and a calibrated variable attenuator to set the input flux of photons to 80,800 /s, we measured the SDE of the devices to be up to 80%, as represented in Fig. 4.(a). We point out that due to the high critical current density ($J_c$) of the material, we also fabricated detectors from slightly thinner films, allowing to boost further the internal detection efficiency at the expense to a reduced critical current, although maintained high, and a reduced yield. With 30% of the detectors made from 9 nm thick films, system detection efficiencies of ~80% at 1550 nm and wide efficiency saturation starting between 0.8 and 0.9 $I_c$ were achieved at 2.5 K. We attribute this remarkable fabrication yield for polycrystalline materials to the use of rather thick films, which are less sensitive to the substrate roughness. To measure the time jitter of the detector, we used a pulsed laser at 1550 nm, a calibrated variable attenuator and a 4 GHz, 40 GSample/s oscilloscope with a combined timing jitter of < 2 ps, and we measured the histogram of the time delay of the coincidence counts between the laser synchronization signal and the SNSPD pulses, as shown in Fig. 4.(b). The fitted setup timing jitter was measured to be as low as 19.47 ps FWHM.

## 4. Conclusions

In summary, we have shown tailored composition and superconducting properties of $Nb_xTi_{1-x}N$

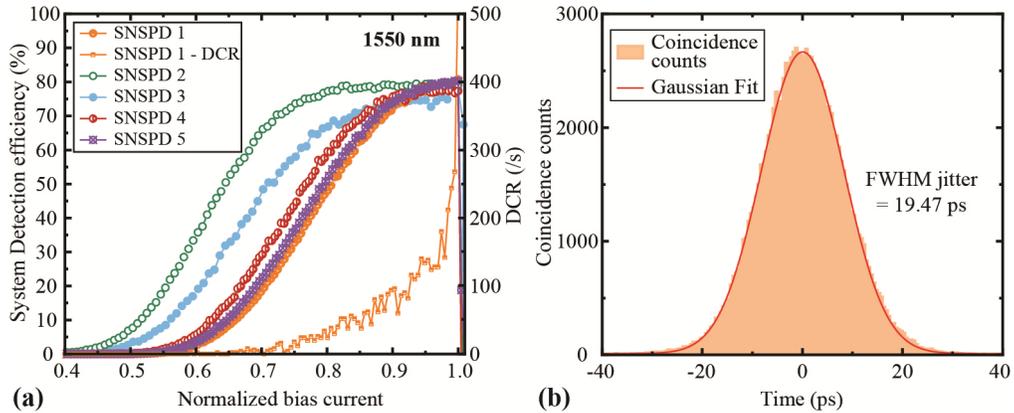

Fig. 4: (a) Measurement at 2.5 K of the System Detection Efficiency at 1550 nm of five detectors and the Dark Count Rate (DCR) of one of the detectors vs. bias current for 20 µm active area devices made on DBR substrates from a 9 nm thick film. (b) Coincidence histogram of the SNSPD on DBR wafer with the laser synchronization signal, revealing a FWHM timing jitter of 19.47 ps.

thin films by controlling the power applied to separate targets in a reactive co-sputtering deposition system. We verified that the influence of the applied powers on the $T_c$ and on the saturation capabilities of the detectors were mainly due to a change in the chemical composition of the ternary alloy. We fabricated devices from 9 nm thick films covering an active area of 20 µm with saturated detection efficiency at 1550 nm at 2.5 K, before moving on to realize fully packaged devices on DBR substrates with 80 % SDE at 1550 nm, a timing jitter of 19.47 ps, and a fabrication yield of 30%. Finally, the minimal impact on the film properties when deposited on SiN suggests the possibility to use this room temperature optimized recipe for the production of on-chip detectors integrated in temperature sensitive photonic circuits [31].

### Acknowledgements

S.S. acknowledges Mukhes Sowwan (Okinawa Institute of Science and Technology OIST, Japan) for access to the transmission electron microscopy facilities. V.Z. acknowledges funding by the VR grant for international recruitment of leading researchers (ref: 2013-7152) and the Knut and Alice Wallenberg Foundation grant "Quantum Sensors". I.E.Z. acknowledges funding from the NWO LIFT-HTSM (680-91-202) grant.